\documentclass[prd,onecolumn]{revtex4}
\usepackage{dcolumn}
\usepackage{multirow}
\usepackage{graphicx}
\usepackage{amssymb}
\usepackage{bm}
\usepackage{hyperref}
\usepackage{epstopdf}
\usepackage{color}
\usepackage{mathrsfs}
\usepackage{amsmath,amssymb,amsthm}
\begin{document}
\title{Traversable braneworld wormholes supported by astrophysical observations}
\author{Deng Wang}
\email{Cstar@mail.nankai.edu.cn}
\affiliation{Theoretical Physics Division, Chern Institute of Mathematics, Nankai University,
Tianjin 300071, China}
\author{Xin-He Meng}
\email{xhm@nankai.edu.cn}
\affiliation{Department of Physics, Nankai University, Tianjin 300071, P.R.China}
\begin{abstract}
In this study, we investigate the characteristics and properties of a traversable wormhole constrained by the current astrophysical observations in the framework of modified theories of gravity (MOG). As a concrete case, we study traversable wormhole space--time configurations in the Dvali--Gabadadze--Porrati (DGP) braneworld scenario, which are supported by the effects of the gravity leakage of extra dimensions. We find that the wormhole space--time structure will open in terms of the $2\sigma$ confidence level when we utilize the joint constraints supernovae (SNe) Ia + observational Hubble parameter data (OHD) + Planck + gravitational wave (GW) and $z<0.2874$. Furthermore, we obtain several model-independent conclusions, such as (i) the exotic matter threading the wormholes can be divided into four classes during the evolutionary processes of the universe based on various energy conditions; (ii) we can offer a strict restriction to the local wormhole space--time structure by  using the current astrophysical observations; and (iii) we can clearly identify a physical gravitational resource for the wormholes supported by astrophysical observations, namely the dark energy components of the universe or equivalent space--time curvature effects from MOG. Moreover, we find that the strong energy condition is always violated at low redshifts.

\end{abstract}
\maketitle
Keywords: braneworld model, traversable wormholes, astrophysical observations.

PACS numbers: 98.80.-k,  95.36.+x, 95.30.Sf.
\section{Introduction}
Almost two decades ago, distance measurements of type Ia supernovae (SNe Ia) indicated that the universe is undergoing a phase of late-time acceleration \cite{1,2}. More recently, this exotic phenomenon remains strongly confirmed by different astrophysical observations, including measurements of baryon acoustic oscillations (BAO) from galaxy surveys, measurements of temperature anisotropies of cosmic microwave background (CMB) radiation, observational Hubble parameter data (OHD), the abundance of galaxy clusters (AGC), and strong and weak gravitational lensing (SGL and WGL). Nonetheless, theoretically, the absence of a reasonable physical mechanism responsible for the late-time accelerated expansion has inspired a great deal of alternatives.

In general, there exist two different approaches to explain the accelerated mechanism in the literature. The first approach is to introduce a new field in the framework of the general theory of relativity (GR), namely, the so-called dark energy. The most straightforward and elegant explanation is the well-known cosmological constant $\Lambda$ model, i.e., the $\Lambda$CDM model, which has been demonstrated to be very successful in depicting many aspects of the observed universe. For example, the late-time acceleration of the universe, the spectrum of anisotropies of the CMB radiation, and the large-scale structure of matter distribution at the linear level are well described by the base cosmology scenario. The newest results of Planck-2015 indicate that there are still some anomalies that are incompatible with the predictions of the base cosmology scenario \cite{3}, specifically, the anomalies of the observed Hubble parameter $H(z)$ and the higher amplitude of the fluctuation spectrum than that inferred from some analyses of rich cluster counts and weak gravitational lensing. Apart from these anomalies, the $\Lambda$CDM scenario also faces the challenge of two fatal problems, i.e., the ``fine-tuning'' problem and the ``coincidence'' problem (see Ref. \cite{4} for details). In addition, Witten proposed that a positive cosmological constant $\Lambda$ is inconsistent with perturbed string theory \cite{5}. Therefore, the base cosmology scenario may not reflect the behaviors of the true universe. Based on this concern, numerous alternative dark energy scenarios have been proposed by theorists in recent years, for instance, phantom \cite{6}, quintessence \cite{7,8,9,10,11,12,13,14}, decaying vacuum \cite{15}, dark fluid \cite{16,17,18,19,20,21}, and Chaplygin gas \cite{22}. The second approach is to introduce modifications to the Einstein--Hilbert action when GR breaks down at substantially large scales, namely the so-called modified theories of gravity (MOG). In general, the popular scenarios belonging to MOG contain $f(R)$ gravity \cite{23,24,25,26,27,28}, scalar--tensor gravity \cite{29,30,31,32}, braneworld gravity \cite{33,34,35}, Einstein--aether gravity \cite{36,37}, and Chern--Simons gravity \cite{38}.

In particular, braneworld gravity is a substantially interesting scenario based on the basic idea that the observational universe could be a (3+1)-dimensional surface (the brane) embedded in a (3+1+d)-dimensional space--time (the bulk), with the Standard Model fields and particles trapped on the brane while gravity can be free to access the bulk. This theory of gravity is inspired by the developments of the $M$ theory \cite{35,a2}. To be more precise, the (9+1)-dimensional superstring theories are encompassed by the (10+1)-dimensional $M$ theory, which is widely considered as a prospective route to quantum gravity. Braneworld scenarios have two main features: (i) at low energies, gravity is confined on the brane and GR is naturally recovered; (ii) at high energies, gravity leaks into the bulk, exhibiting behaviors in a high-dimensional way. The gravitational effects of extra dimensions offer attractively testable implications for high-energy astrophysics and cosmology, which result from the $M$ theory. More concretely, the accelerated universe could be the result of gravitational leakage into extra dimensions over Hubble distances rather than the consequence of a nonzero cosmological constant \cite{a3}. In the present situation, we just take into account the simplest braneworld scenario where the four-dimensional gravity on the brane is modified at low energies, becoming five-dimensional Dvali--Gabadadze--Porrati (DGP) scenarios, which is one class of the five-dimensional braneworld scenarios based on the Randall--Sundrum (RS) scenarios.

As in our previous works \cite{39,40,41,42}, the present study is aimed at exploring the astrophysical scale properties of the braneworld scenario by assuming that the dark energy fluid permeates over the whole universe. More precisely, we investigate the properties and features of a mysterious astrophysical object, i.e., a wormhole, which has attracted considerable attention today along with other astrophysical objects, such as pulsars, white dwarfs, and black holes.

Wormholes can be defined as a class of special space--time structure connecting two different universes or two widely separated regions of our own universe. In recent years, gradually mounting interest in the subject is mainly ascribed to the stirring and elegant discovery that the universe is experiencing a phase of accelerated expansion. Because the null energy conditions (NECs) are violated in both cases, an amazing and unexpected overlap occurs between these two seemingly separated subjects. More recently, as demonstrated in our previous works \cite{39,40,41,42}, we are dedicated to investigating the related properties and features of wormhole space--time structure for a given dynamical dark energy scenario. In the present study, we explore traversable wormholes in the framework of MOG. As a concrete case, traversable wormholes supported by astrophysical observations in the DGP scenario are investigated. In what follows, it is very necessary to provide a brief historical review on wormholes.

The first study of wormhole physics can be traced back to Flamm in 1916, when he analyzed the newly discovered Schwarzschild solution \cite{43}. Subsequently, in 1935, Einstein and Rosen (ER) proposed the so-called ER bridge when they attempted to construct a geometrical model for a physical fundamental particle, e.g., an electron \cite{44}. After the pioneering study by ER, this field lay dormant for almost two decades. Then, in 1955, Wheeler developed the appealing concept of a ``geon, '' which was assumed to be the solution of the coupled Einstein--Maxwell equation \cite{45}. From the point of view of topology, he considered a multiply-connected space--time structure, where two distant regions were linked by a tunnel, namely, the geon. Furthermore, Misner and Wheeler conducted a series of research projects on differential geometry and abstract topology in physics. In their 1957 paper \cite{46}, they introduced the term ``wormhole'' for the first time. During the following 30 years, the wormhole field lay dormant once again, except for Bronnikov's tunnel-like solutions \cite{47} and Ellis's drainhole \cite{48,49}. In 1988, a new era of wormhole physics was opened by Morris and Thorne in light of their milestone paper \cite{50}, where they analyzed in detail the construction of the wormhole, energy conditions, time machines, the stability problem, and traversabilities of wormholes. In succession, Visser and Possion introduced the interesting ``thin-shell wormhole'' by conjecturing that all the ``exotic matter'' is confined to a thin shell between universes \cite{51,52,53,54}. Because of the discovery of the late-time acceleration of the universe in 1998, a great deal of related research about wormholes has been motivated by developments of various kinds of phantom-like dynamical dark energy scenarios or MOG scenarios \cite{55,56}. Subsequently, for the first time, we investigated traversable wormholes constrained by modern astrophysical observations \cite{57,58,59}. This is also the starting point of our present work.

This paper is organized as follows. In the next section, we present a brief review of wormholes. In Section 3, we also review the DGP braneworld scenario briefly and constrain the DGP scenario by using the current astrophysical observations to investigate the wormhole space--time configuration quantitatively. In Section 4, two specific traversable wormhole solutions are presented and the related characteristics are also studied.  In Section 5, we explore the energy conditions based on the current astrophysical observations on both cosmological and astrophysical scales. Our discussion and conclusions are presented in the final section. Note that throughout the paper we use the units $8\pi G=c=1$.

\section{Brief review of wormholes}
Consider a static, spherical symmetric metric for the wormhole space--time configuration,
\begin{equation}
ds^2=-e^{2\Phi(r)}dt^2+\frac{dr^2}{1-b(r)/r}+r^2(d\theta^2+\sin^2\theta d\phi^2), \label{1}
\end{equation}
where $r$ is the radial coordinate running in the range [$r_0$, $\infty$) (where $r_0$ denotes the throat radius of a wormhole) and $\theta$ and $\phi$ represent the angular coordinates. $\Phi(r)$ denotes the redshift function since it is related to the gravitational redshift, and $b(r)$ denotes the shape function, for it can determine the spatial shape of the wormhole.

As described in our previous works \cite{39,40,41,42}, in general, four fundamental ingredients are required to form a traversable wormhole:

$\star$ Violate the NEC, i.e., $T_{\mu\nu}k^{\mu}k^{\nu}>0$, where $T_{\mu\nu}$ is the stress--energy tensor and $k^{\mu}$ is any future directed null vector.

$\star$ Satisfy the flare-out conditions, i.e., $b(r_0)=r_0$, $b'(r_0)<1$ and $b(r)<r$ when $r>r_0$.

$\star$ To avoid a horizon, the redshift function $\Phi(r)$ should be finite everywhere.

$\star$ Impose asymptotically flatness conditions, i.e., $b/r\rightarrow0$ and $\Phi\rightarrow0$ when $r\rightarrow\infty$.

Using the Einstein field equation $G_{\mu\nu}=T_{\mu\nu}$ in an orthonormal reference frame, we can obtain the stress--energy scenario as follows:
\begin{equation}
\rho=\frac{b'}{r^2}, \label{2}
\end{equation}
\begin{equation}
p_r=\frac{b}{r^3}-2\frac{\Phi'}{r}\left(1-\frac{b}{r}\right), \label{3}
\end{equation}
\begin{equation}
p_t=\left(1-\frac{b}{r}\right)\left[\Phi''+(\Phi')^2-\Phi'\frac{b'r-b}{2r^2(1-b/r)}-\frac{b'r-b}{2r^3(1-b/r)}+\frac{\Phi'}{r}\right], \label{4}
\end{equation}
where $\rho(r)$ is the matter energy density, $p_r(r)$ is the radial pressure, $p_t(r)$ is the lateral pressure orthogonal to the radial direction, and the prime denotes the derivative with respect to $r$. Using the stress--energy conservation equation, $T^{\mu\nu}_{\hspace{3mm};\nu} = 0$, we obtain
\begin{equation}
p'_r=\frac{2}{r}(p_t-p_r)-\Phi'(\rho+p_r), \label{5}
\end{equation}
which can be regarded as the hydrostatic equation of equilibrium for the materials supporting a wormhole or the relativistic Euler equation. In the following section, we review the DGP braneworld scenario briefly to constrain it by cosmic observations and study the corresponding traversable wormhole space--time configurations.

\section{The DGP braneworld scenario}
The DGP braneworld scenario, which was first generalized to the cosmology field by Deffayet \cite{a3}, modifies GR at low energies. This scenario produces the self-acceleration of the late-time universe owing to a weakening of gravity at low energies. Similar to the RS scenario, the DGP scenario is also a five-dimensional scenario with infinite extra dimensions. The dynamics of gravity is governed by a competition between a Ricci scalar term in the four-dimensional brane and
an Einstein--Hilbert action in the five-dimensional bulk. Then, the first Friedmann equation of the DGP scenario can be expressed as \cite{60}
\begin{equation}
H^2=H_0^2\{\Omega_{k0}(1+z)^2+[\sqrt{\Omega_{r_c}}+\sqrt{\Omega_{r_c}+\Omega_{m0}(1+z)^3}]^2\}, \label{6}
\end{equation}
where $H$ denotes the Hubble parameter, $H_0$ is the present value of the Hubble parameter, $\Omega_{m0}$ and $\Omega_{k0}$ are the fractional contributions of matter and curvature, respectively, and $\Omega_{r_c}=1/4r_c^2H_0^2$ is the bulk-induced term with respect to the crossover radius $r_c$. The $z=0$ value of Eq. (\ref{6}) leads to the usual normalization condition $\Omega_{k0}+(\sqrt{\Omega_{r_c}}+\sqrt{\Omega_{r_c}+\Omega_{m0}})^2=1$, and for a flat universe ($\Omega_{k0}=0$), $\Omega_{r_c}=(1-\Omega_{m0})^2/4$. Furthermore, Eq. (\ref{6}) can be rewritten as
\begin{equation}
E(z)=\{[1-(\sqrt{\Omega_{r_c}}+\sqrt{\Omega_{r_c}+\Omega_{m0}})^2](1+z)^2+[\sqrt{\Omega_{r_c}}+\sqrt{\Omega_{r_c}+\Omega_{m0}(1+z)^3}]^2\}^{1/2}, \label{7}
\end{equation}
where $E(z)$ denotes the normalized Hubble parameter. Furthermore, according to Ref. \cite{60}, the $z$-dependent equation of state (EoS) of the DGP scenario can be expressed as
\begin{equation}
\omega_{eff}(z)=\{[\sqrt{4\Omega_{r_c}/\Omega_{m0}(1+z)^3+4}][\sqrt{\Omega_{r_c}/\Omega_{m0}(1+z)^3}+\sqrt{\Omega_{r_c}/\Omega_{m0}(1+z)^3+1}]\}^{-1}-1. \label{8}
\end{equation}
In what follows, we exhibit our methodology to constrain the DGP scenario by using the current astrophysical observations.
\begin{table}[h!]
\begin{tabular}{ccccccc}
\hline
                           &SNe Ia+OHD+GW   & SNe Ia+OHD+Planck-2015+GW\\
\hline
$\chi^2_{min}$             &$579.444$        &$579.455$\\
$\Omega_{m0}$                   &$0.2660^{+0.0553+0.0927}_{-0.0524-0.0858}$       &$0.2684^{+0.0552+0.0903}_{-0.0522-0.0864}$\\
$\Omega_{r_c}$                    &$0.1902^{+0.0221+0.0369}_{-0.0209-0.0346}$       &$0.1912^{+0.0229+0.0379}_{-0.0203-0.0347}$\\
\hline
\end{tabular}
\caption{Minimum values of the derived $\chi^2$; corresponding $1\sigma$, $2\sigma$, and $3\sigma$ confidence intervals; and the best-fitting values of the model parameter pair ($\Omega_{m0}$, $\Omega_{r_c}$) in the DGP braneworld scenario obtained by using different kinds of constraints: SNe Ia, OHD, Planck-2015, and GW.}
\label{tab1}
\end{table}

In the present work, we utilize the Union 2.1 data sets for numerical analysis; these are composed of 580 SNe Ia data points covering the redshift range [0.015, 1.4].  To exhibit the standard $\chi^2$ statistical estimates, the theoretical distance modulus is defined as $\mu_{th}(z_i)=m-M=5\log_{10}d_L(z_i)+25$, where $m$ is the apparent magnitude, $M$ is the absolute magnitude, and $d_L(z_i)$ is the luminosity distance at a given redshift $z_i$ in units of Mpc. The luminosity distance is defined as $d_L(z_i)=(1+z_i)\int^{z_i}_0\frac{dz'}{E(z';\theta)}$, where $\theta$ represents a set of model parameters. Then, the corresponding $\chi^2$ for the SNe Ia observations is written as $\chi^2_{SN}=\sum^{580}_{i=1}[\frac{\mu_{obs}(z_i)-\mu_{th}(z_i;\theta)}{\sigma_i}]^2$, where $\mu_{obs}$ is the observed value of the distance modulus and $\sigma_i$ is the corresponding 1$\sigma$ statistical error at a given redshift $z_i$. We use the latest OHD set consisting of 36 data points, which are obtained from model-independent observations and avoid integrating over the redshift $z$. The corresponding $\chi^2$ for the OHD is defined as $\chi^2_{OHD}=\sum^{36}_{i=1}[\frac{H_0E(z_i)-H_{obs}(z_i)}{\sigma_i}]^2$, where $H_{obs}(z_i)$ is the observed value of the Hubble parameter at a given redshift $z_i$. Another important and powerful constraint comes from the CMB anisotropy observations, and we use the Planck-2015 data sets to constrain the DGP braneworld scenario. For simplicity, we take the shift parameter $\mathcal{R}$ instead of the full data from the CMB anisotropy observations, because taking the entire CMB data to perform a global fitting will expend a great deal of power and computation time. The shift parameter $\mathcal{R}$ is defined as $\mathcal{R}=\sqrt{\Omega_{m0}}\int^{z_C}_0\frac{dz'}{E(z')}$, where $z_C$ is the redshift of recombination and $\Omega_{m0}$ is the present-day value of the matter density ratio parameter. As shown in a recent paper \cite{3}, the shift parameter for the Planck-2015 data sets is $\mathcal{R}=1.7482\pm0.0048$ and $z_C=1089.90$. The corresponding $\chi^2$ for the CMB anisotropy observations can be expressed as $\chi^2_{CMB}(\theta)=[\frac{\mathcal{R}_{obs}-\mathcal{R}(\theta)}{\sigma_{\mathcal{R}}}]^2$, where $\mathcal{R}_{obs}$ and $\sigma_{\mathcal{R}}$ denote the values of the shift parameter and the corresponding $1\sigma$ statistical error, respectively. In addition, as in our previous works \cite{57,58,59}, we still take a single data point from the single gravitational-wave (GW) event GW150914 \cite{61}. Note that, although the quality of the current gravitational-wave data is not very good, we believe that the gradually mounting data in the future will place tighter constraints on various kinds of cosmological scenarios. To be more precise, we use the luminosity distance $420^{+150}_{-180}$ Mpc of two supermassive black holes as a complementary probe. The corresponding $\chi^2$ for GW150914 is expressed as $\chi^2_{GW}$. Naturally, the expected $\chi^2$ of the combined constraints from the SNe Ia, OHD, CMB and GW150914 data sets can be defined as
\begin{equation}
\tilde{\chi}^2={\chi}^2_{SN}+{\chi}^2_{OHD}+\chi^2_{CMB}++\chi^2_{GW} \label{9}.
\end{equation}

\begin{figure}
\centering
\includegraphics[scale=0.5]{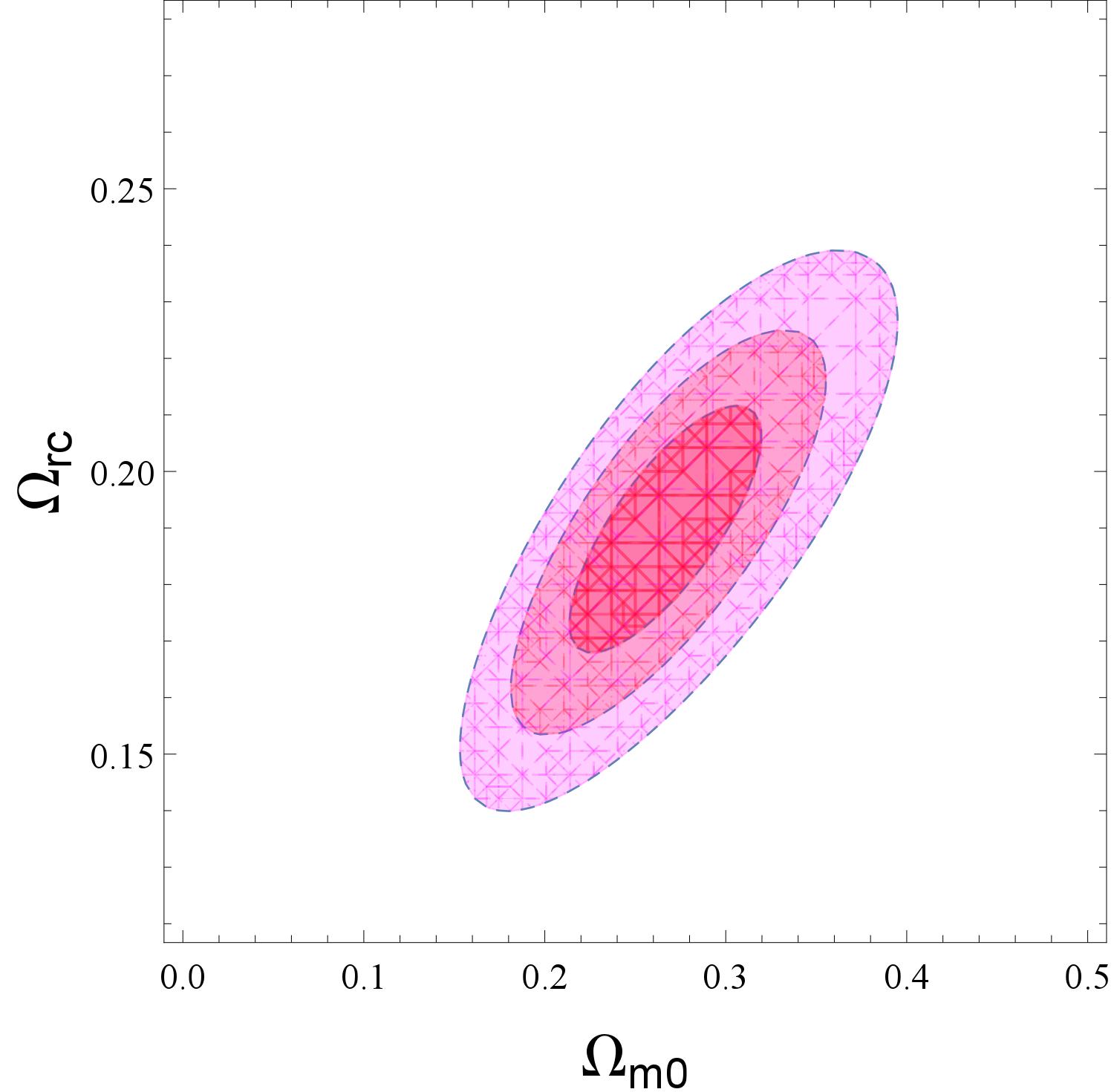}
\caption{$1\sigma$, $2\sigma$, and $3\sigma$ confidence ranges for the model parameter pair ($\Omega_{m0}$, $\Omega_{r_c}$) of the DGP braneworld scenario constrained by SNe Ia+OHD+GW data sets.}\label{f1}
\end{figure}
\begin{figure}
\centering
\includegraphics[scale=0.5]{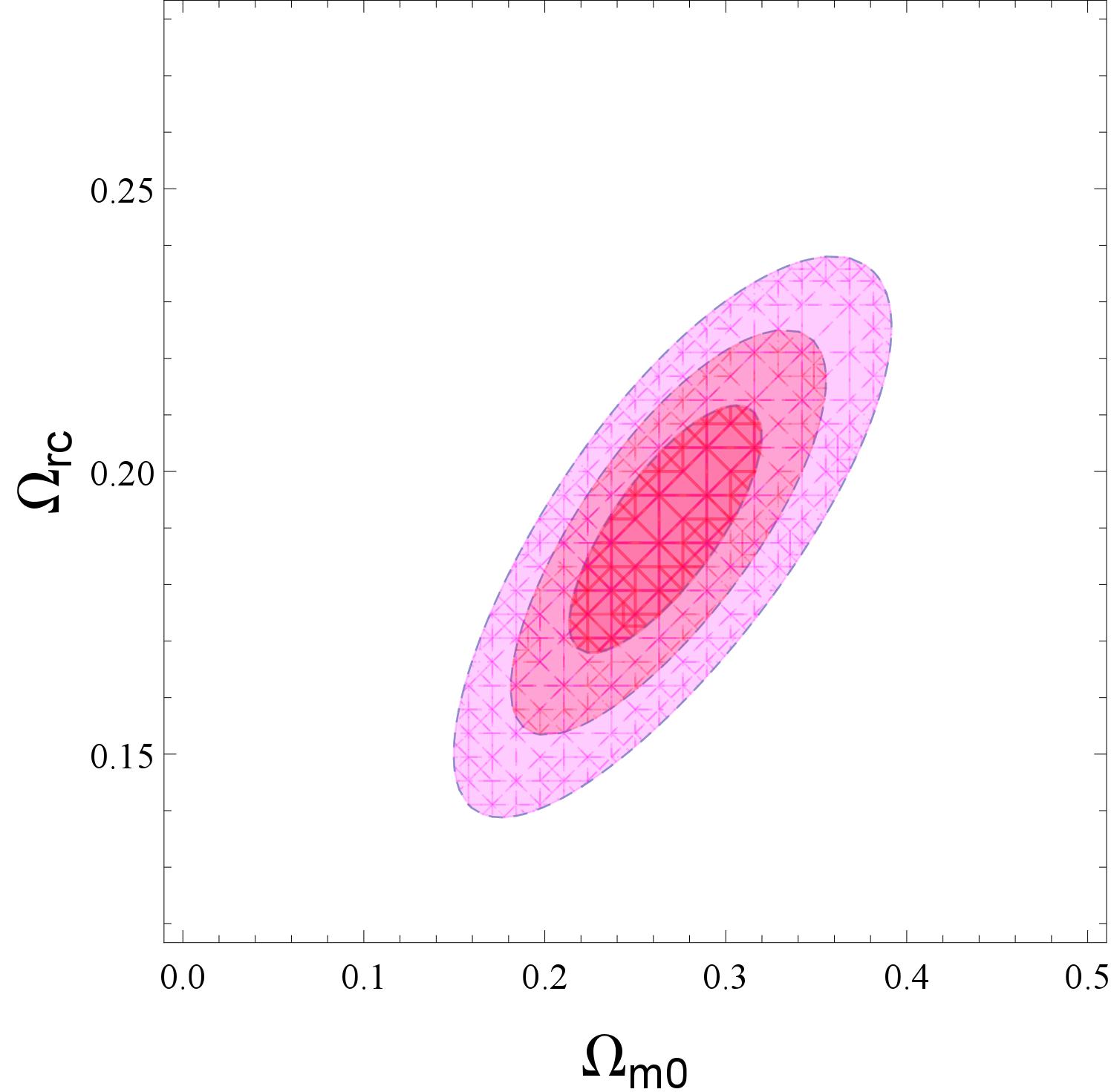}
\caption{$1\sigma$, $2\sigma$, and $3\sigma$ confidence ranges for the model parameter pair ($\Omega_{m0}$, $\Omega_{r_c}$) of the DGP braneworld scenario constrained by SNe Ia+OHD+Planck-2015+GW data sets.}\label{f2}
\end{figure}
\begin{figure}
\centering
\includegraphics[scale=0.5]{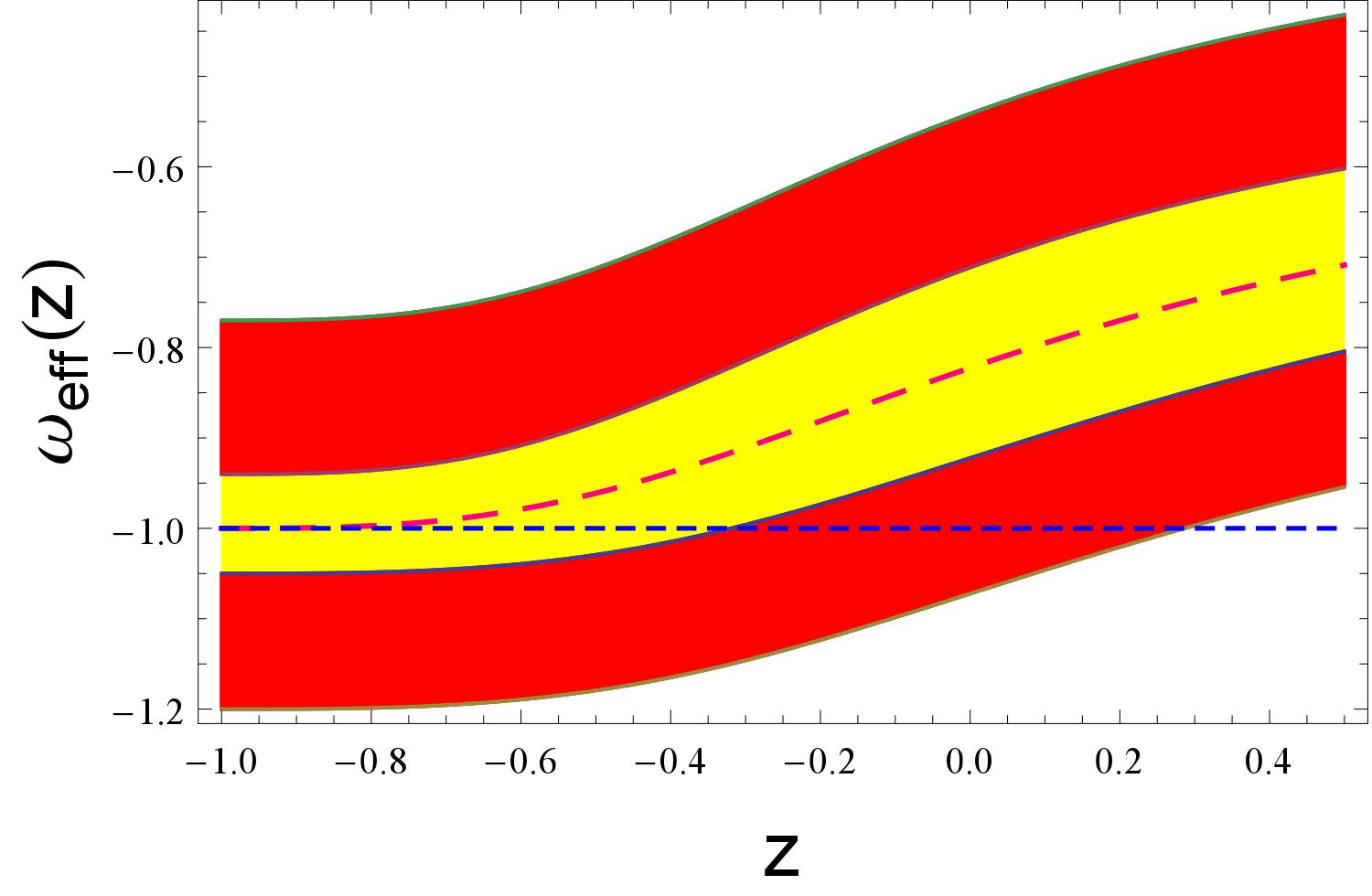}
\caption{Evolutionary tendency of the EoS of the DGP braneworld scenario constrained by SNe Ia+OHD+Planck+GW data sets. The long-dashed (red) line represents the EoS when we take the best-fitting values of the model parameter pair ($\Omega_{m0}$, $\Omega_{r_c}$), and the short-dashed (blue) line represents the base cosmology scenario. The yellow region corresponds to the $1\sigma$ confidence band, and the red regions reflect the parts where the $2\sigma$ confidence band surpasses the $1\sigma$ confidence band.}\label{f3}
\end{figure}
\begin{figure}
\centering
\includegraphics[scale=0.5]{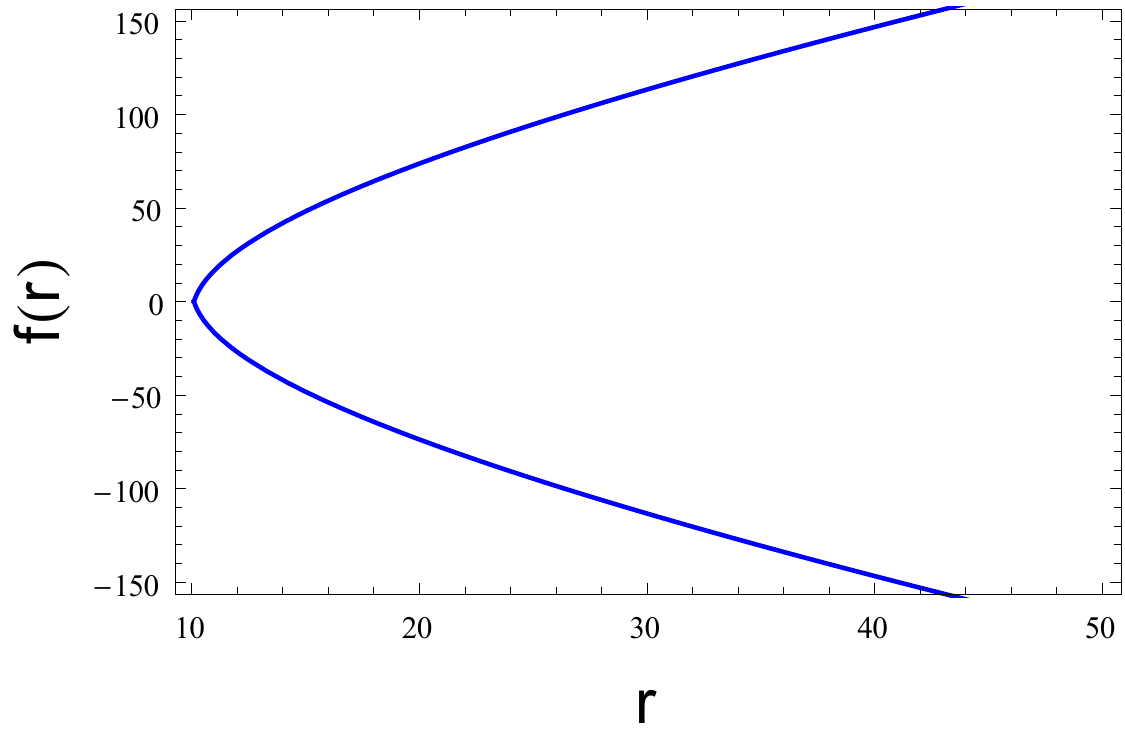}
\caption{Profile curve of a traversable wormhole.}\label{f4}
\end{figure}
\begin{figure}
\centering
\includegraphics[scale=0.5]{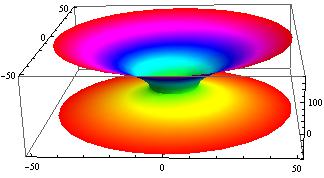}
\caption{Embedding diagram generated by rotating the profile curve about the vertical axis.}\label{f5}
\end{figure}

The minimum values of the derived $\chi^2$ for two different combined constraints, the best fitting values of the model parameter pair ($\Omega_{m0}$, $\Omega_{r_c}$), and the corresponding $1\sigma$ and $2\sigma$ confidence intervals are listed in Table \ref{tab1}. Furthermore, the likelihood distributions of two different constraints for the DGP scenario are depicted in Figs. \ref{f1} and \ref{f2}. It is not difficult to find that, in our situation, the results of two different constraints share a high degeneracy; we cannot produce a tighter constraint apparently when adding Planck-2015 data to the joint constraints SNe Ia+OHD+GW.

\section{Specific traversable wormholes}
We explore the related properties of traversable wormholes of the DGP scenario in this section. Through some simple calculations, we find that the EoS parameter $\omega_{eff}(z)<-1$ in terms of the $2\sigma$ confidence level when $z<0.2874$ (the value at which wormholes appear); i.e., the NEC is violated by the effects of the gravity leakage of the extra dimensions in the DGP braneworld scenario. It is worth noting that, in our previous works, we just considered the EoS when the model parameters take the best-fitting values, which is insufficient for providing complete information on the survival range of the wormholes for a given cosmological scenario. In addition, the wormholes may not come into being immediately when the EoS parameter $\omega_{eff}(z)<-1$ and the specific formation mechanism of wormholes still needs to be explored further.
\subsection{ Constant redshift function}
Considering the simplest case $\Phi=C$, where $C$ is an arbitrary constant, we can obtain the shape function by replacing Eqs. (\ref{2}) and (\ref{8}) in Eq. (\ref{3}):
\begin{equation}
b(r)=r_0^{\frac{1}{\omega_{eff}}+1}r^{-\frac{1}{\omega_{eff}}}. \label{10}
\end{equation}
It is not difficult to find that the flare-out conditions are well satisfied by numerical calculations. To be more precise, the shape function $b(r)<r$ when $r>r_0$ and evaluating at the throat $r_0$, we can obtain
\begin{equation}
b'(r_0)=-\frac{1}{\omega_{eff}}. \label{11}
\end{equation}
Subsequently, if taking the model parameter pair ($\Omega_{m0}$, $\Omega_{r_c}$) from the joint constraints of SNe Ia+OHD+Planck-2015+GW data sets in terms of the lower boundary of the $2\sigma$ confidence band (see Fig. \ref{f3}), we can obtain $b'(r_0)=0.931099<1$ ($\omega_{eff}=-1.074$) at the present epoch. The solution is both traversable and asymptotically flat, because the redshift function $\Phi$ is finite everywhere and $b/r\rightarrow0$ when $r\rightarrow\infty$. Furthermore, it is constructive and necessary to depict the space--time structure of the traversable wormhole. As described in Ref. \cite{50}, we can obtain the wormhole space--time configuration through the function $f(r)$, which characterizes the embedded surface of the wormhole. This function is defined as
\begin{equation}
\frac{df}{dr}=\pm\frac{1}{\sqrt{\frac{r}{b(r)}-1}}. \label{12}
\end{equation}
Then, we can construct the profile curve and the concrete embedded diagram by solving this equation numerically. It is noteworthy that we assumed a throat radius of the wormhole of $r_0=10$ m and an EoS parameter of $\omega_{eff}=-1.074$ (see Figs. \ref{f4} and \ref{f5}). Following our previous studies \cite{39,40,41,42}, here we conduct a wormhole traversability analysis, which may be the most interesting and attractive consideration in wormhole physics. In general, three necessary ingredients are required for a human being in a spaceship who tries to successfully traverse a wormhole:

$\star$ The acceleration felt by the travelers should not exceed Earth's gravitational acceleration $g_\oplus$.

$\star$ The tidal acceleration should not exceed Earth's gravitational acceleration.

$\star$ The traversal time measured by the travelers and the observers who stay at rest at a space station should satisfy several quantitative relationships.

For the conveniences of calculations, we just derive the key formula obtained by some straightforward calculations in the following manner:
\begin{equation}
v\leqslant r_0\sqrt{\frac{\omega_{eff}(z) g_\oplus}{\omega_{eff}(z)+1}}, \label{13}
\end{equation}
where $v$ is the traversal velocity, $r_0$ is the throat radius of the traversable wormhole, and $g_\oplus$ is Earth's gravitational acceleration. We can easily find that the traversal velocity depends on the redshift for the DGP scenario, which also implies that the wormhole space--time configuration evolves with cosmic time. As a concrete case, assuming $r_0=100$ m and taking the best-fitting values (0.2684, 0.1912) of the model parameter pair ($\Omega_{m0}$, $\Omega_{r_c}$) from the SNe Ia+OHD+Planck+GW constraints, we obtain a traversal velocity of $v\approx1192.6$ m/s at the present epoch. In what follows, choosing the matching radius $D=10000$ m, we can also obtain a traversal time of $\Delta\tau\approx\Delta t\approx2L/v\thickapprox16.77\;\rm s$.

\subsection{A special shape function: $b(r)=r_0+\frac{1}{\omega_{eff}}(r_0-r)$}
Considering a specific choice for the shape function $b(r)=r_0+\frac{1}{\omega_{eff}}(r_0-r)$ and utilizing Eq. (\ref{8}), we can get
\begin{equation}
\Phi'(r)=-\frac{1}{2r}. \label{14}
\end{equation}
Integrating over $r$ on both sides gives
\begin{equation}
\Phi(r)=-\frac{1}{2}\ln r+B, \label{15}
\end{equation}
where $B$ is an integration constant. We can easily find that the solution is not asymptotically flat because it diverges when $r\rightarrow\infty$. Thus, the wormhole space--time configuration is not traversable for an interstellar traveler. However, in theory, we can construct a traversable wormhole by gluing an exterior flat space--time geometry onto the interior space--time geometry at a junction radius $r_m$. Hence, the integration constant $B$ can be expressed as $C=\Phi(r_m)+\frac{1}{2}\ln(\frac{r_m}{r})$. Furthermore, we can calculate the amounts of the exotic matter threading the wormhole by using the ``volume integral quantifier '' (VIQ) method. It is noteworthy that the amounts of  exotic matter can be described effectively by the definite integral $\int T_{\mu\nu}k^\mu k^\nu dV$ and that here we just consider the finite range of the traversable wormhole, $r\in[r_0, r_m]$. Subsequently, using the corresponding quantity $I_V=\int[p_r(r)+\rho]dV$, we can obtain the following relation:
\begin{equation}
I_V=\int^{r_m}_{r_0}(r-b)\left[\ln\left(\frac{e^{2\Phi}}{1-b/r}\right)\right]'dr. \label{16}
\end{equation}
It follows that
\begin{equation}
I_V=\frac{[1+\omega_{eff}(z)](r_0-r_m)}{\omega_{eff}(z)}. \label{17}
\end{equation}
Obviously, this physical quantity is sensitive to the redshift $z$. Like the case of the constant redshift function, we also take the best-fitting values (0.2684, 0.1912) of the model parameter pair ($\Omega_{m0}$, $\Omega_{r_c}$) from the SNe Ia+OHD+Planck+GW constraints and obtain
\begin{equation}
I_V=0.0689013(r_0-r_m). \label{18}
\end{equation}
It is not difficult to verify that the physical quantity $I_V\rightarrow0$ when $r_m\rightarrow r_0$, which means that, in theory, we can construct a traversable wormhole with infinitesimal amounts of averaged-NEC-violating dark energy fluid in the DGP braneworld scenario.
\begin{figure}
\centering
\includegraphics[scale=0.5]{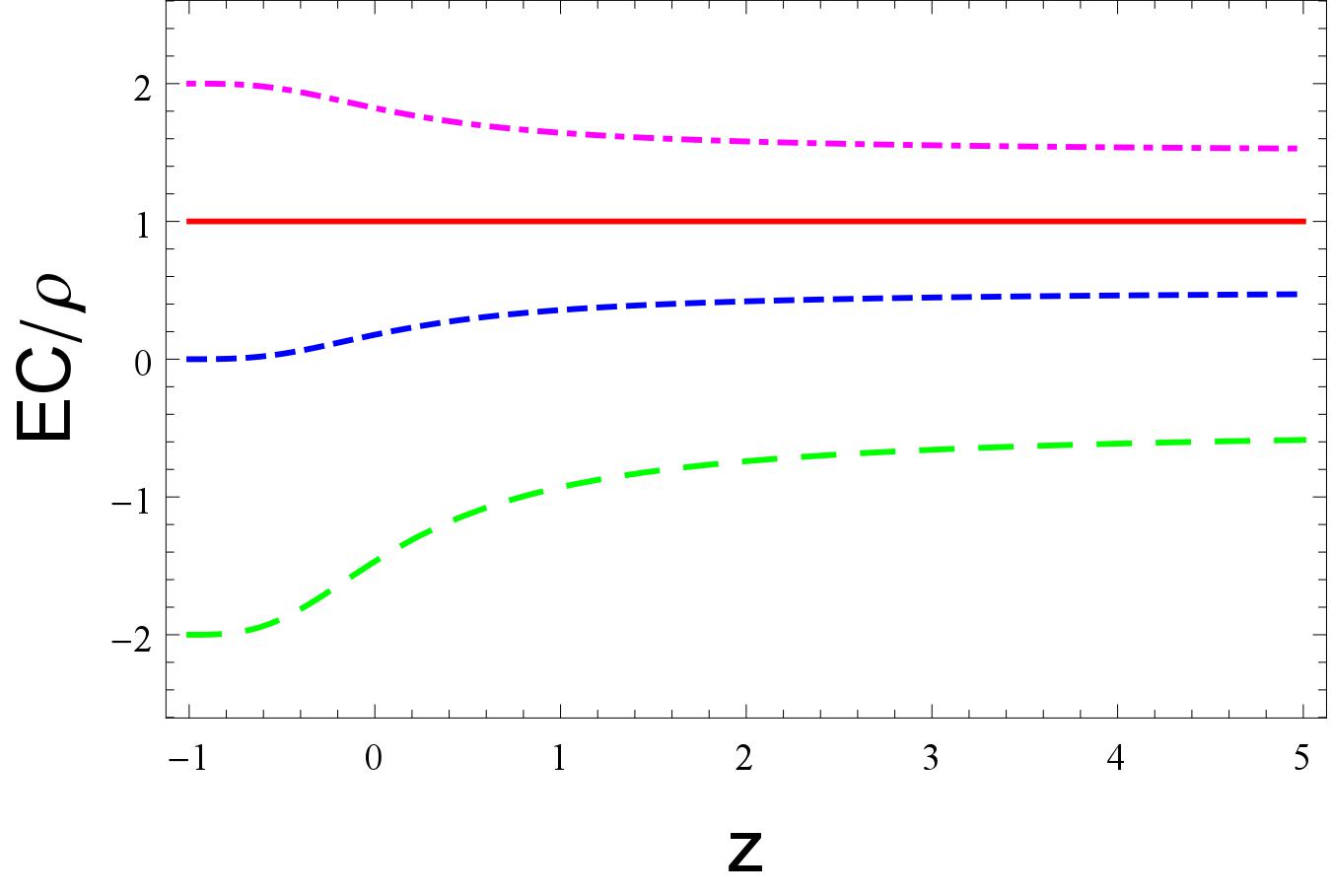}
\caption{Relations between different energy conditions and the wormhole radius $r$. The green (long-dashed) line represents the expression $(p+3\rho)/\rho=3\omega_{eff}+1$, the blue (short-dashed) line represents the expression $(p+\rho)/\rho=\omega_{eff}+1$, the pink (dot-dashed) line represents the expression $(\rho-p)/\rho=1-\omega_{eff}$, and the red (solid) line represents the expression $\rho/\rho=1$.}\label{f6}
\end{figure}
\begin{figure}
\centering
\includegraphics[scale=0.5]{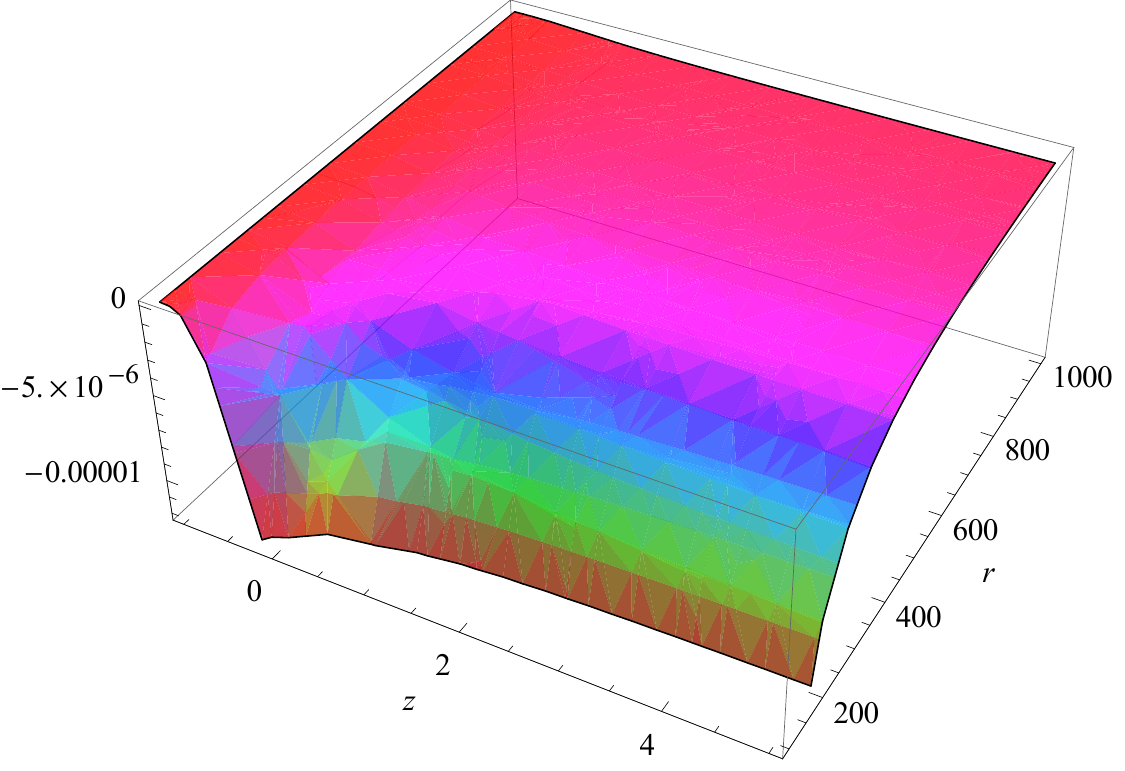}
\includegraphics[scale=0.5]{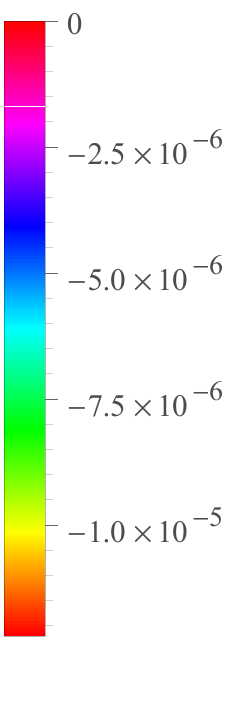}
\caption{ $p+\rho$ as a function of the redshift $z$ and the wormhole radius $r$.}\label{f7}
\end{figure}
\begin{figure}
\centering
\includegraphics[scale=0.5]{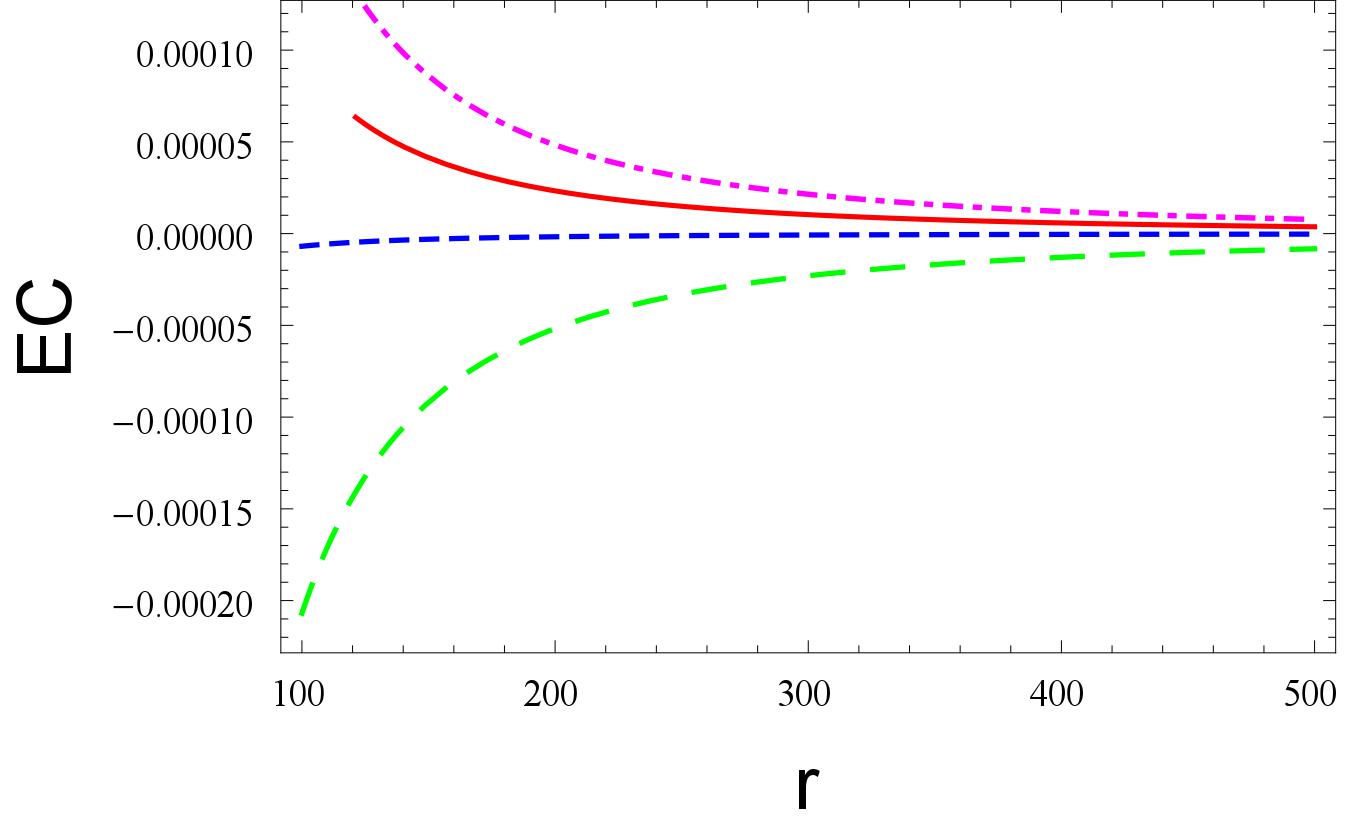}
\caption{Relations between different energy conditions and the wormhole radius $r$. The green (long-dashed) line represents the expression $p+3\rho$, the blue (short-dashed) line represents the expression $p+\rho$, the pink (dot-dashed) line represents the expression $\rho-p$, and the red (solid) line represents the expression $\rho$.}\label{f8}
\end{figure}

\section{Energy conditions}
In general, there are two important types of energy conditions in classical general relativity, i.e., pointwise energy conditions and averaged energy conditions. Generally, the standard pointwise energy conditions contain the above-mentioned NEC, the weak energy condition (WEC), the strong energy condition (SEC), and the dominant energy condition (DEC). The mathematical description of different energy conditions in a spatially flat Friedmann--Robertson--Walker (FRW) universe can be expressed as follows:
\begin{equation}
{\rm NEC} \Leftrightarrow \rho+p\geqslant0, \label{19}
\end{equation}
\begin{equation}
{\rm WEC} \Leftrightarrow \rho+p\geqslant0\quad {\rm and} \quad\rho\geqslant0, \label{20}
\end{equation}
\begin{equation}
{\rm SEC} \Leftrightarrow \rho+p\geqslant0\quad {\rm and} \quad\rho+3p\geqslant0, \label{21}
\end{equation}
and
\begin{equation}
{\rm DEC} \Leftrightarrow \rho\pm p\geqslant0\quad {\rm and} \quad\rho\geqslant0. \label{22}
\end{equation}
It is worth noticing that the NECs are the necessary conditions of the other three energy conditions mathematically.

In the present work, we are dedicated to explore the relationships between the structure formation of traversable wormholes and energy conditions from the point of view of cosmology. Furthermore, we exhibit these underlying relations by using the DGP braneworld scenario in the framework of MOG. Substituting Eq. (\ref{8}) into different conditional inequalities (\cite{19,20,21,22}), we find that the model parameter ($\alpha$, $\beta$) and redshift $z$ will enter the energy conditions and affect the structure formation of traversable wormholes. More precisely, the WEC and DEC are automatically satisfied in the possible range [0.2874, \~{\it z}], where \~{\it z} represents the possible value of the low redshift (see Fig. \ref{f6}). Note that we investigated the energy conditions in terms of the $2\sigma$ confidence level. In addition, we can easily find that the second inequality $\rho+3p\geqslant0$ of the SEC is always violated at low redshifts.

More physically, the DEC implies that the dark energy fluid cannot travel faster than the speed of light and that the nonviolation of the WEC indicates that the dark energy fluid in the DGP scenario has a nondiverging effect on congruences of null geodesics. Meanwhile, the SEC provides a tighter limitation than the WEC but it is still physically reasonable for the stress--energy tensor. In the DGP scenario, we can find that the pressure of dark energy from the gravity leakage of the extra dimensions is large enough at low redshifts. Therefore, the SEC is always violated until the end of the universe.

In what follows, it is very necessary and constructive to study the behaviors of different energy conditions on the astrophysical scale, namely the effects of different energy conditions on wormhole space--time configurations. Here, we just take into account the case of a constant redshift function, using the EoS parameter from Eq. (\ref{8}) and Eq. (\ref{11}), and we can express conveniently the NEC in the following manner:
\begin{equation}
p+\rho=-\frac{1}{r^2}\left(\frac{1}{\omega_{eff}(z)}+1\right)\geqslant0. \label{23}
\end{equation}
In Fig. \ref{f7}, we could find that the NEC violation is closely related to the dimension of the traversable wormhole and that the NEC violation occurs at a very small order of magnitude at low redshifts ($\sim$$10^{-6}$). To understand the size-dependent property of the traversable wormhole better, we depict the size-dependent behaviors of different energy conditions in Fig. \ref{f8} at the present epoch (i.e., $\omega_{eff}(0)=-1.074$). It is easy to find that the behaviors of different energy conditions on the astrophysical scale is compatible with those on the cosmological scale. Moreover, with the gradually increasing radius of the traversable wormhole, the behaviors of different energy conditions in the DGP braneworld scenario will tend to be 0, which is completely consistent with those in the standard cosmological scenario ($p+\rho=0$). In fact, it is substantially reasonable because the energy density $\rho$ of the material threading the wormhole will approach 0 very fast with increasing radius ($\rho=-\frac{1}{\omega_{eff}(z) r^2}$).

\section{Discussion and conclusions}
In this work, for the first time, we investigated the characteristics and properties of a traversable wormhole constrained by the current astrophysical observations in the framework of MOG. As a concrete case, we studied traversable wormhole space--time configurations in the DGP braneworld scenario, which are supported by the effects of the gravity leakage of extra dimensions. As in our previous works \cite{39,40,41,42}, we first constrain the DGP braneworld scenario by using various data sets, including SNe Ia, OHD, Planck, and a single data point from the single event GW150914. We find that the two different joint constraints, i.e., SNe Ia+OHD+GW and SNe Ia+OHD+Planck+GW, share a substantially high degeneracy and that the wormhole space--time structure will open in terms of the $2\sigma$ confidence level when we take the joint constraints SNe Ia+OHD+Planck+GW and $z<0.2874$ (see Fig. \ref{f3}). We then obtain two special traversable wormhole solutions supported by the current astrophysical observations: the case of a constant redshift function and that of a specific choice for the shape function. In what follows, utilizing the model parameter pair from the SNe Ia+OHD+Planck+GW constraint, we analyze the traversabilities of the first traversable wormhole solution and the amounts of exotic matter of the second solution.

In contrast to our previous studies \cite{16,17,18,19}, we explored the energy conditions from the point of view of observational cosmology. More precisely, we obtained several model-independent conclusions: (i) According to the violations of different energy conditions, the exotic matter threading the traversable wormhole can be divided into four classes during the evolutionary processes of the universe. (ii) We can impose a strong restriction to the local wormhole space--time structure by using the current astrophysical observations. (iii) A physically gravitational resource for the wormholes supported by astrophysical observations has been identified, i.e., the dark energy of the universe or equivalent space--time curvature effects from the framework of MOG.

We investigated the energy conditions on both astrophysical and cosmological scales and derived the underlying relations between energy conditions and wormhole structure formation for the DGP braneworld scenario from the point of view of observational cosmology. More specifically, we found that the WEC and the DEC are automatically satisfied in the possible range [0.2874, \~{\it z}] (see Fig. \ref{f6}). Moreover, we can easily find that the second inequality $\rho+3p\geqslant0$ of the SEC is always violated at low redshifts. Furthermore, we also interpret clearly the physical implications of violations of different energy conditions for the DGP braneworld scenario.

With gradually increasing amounts of astrophysical data, we expect to test more cosmological scenarios, study the large-scale structure more thoroughly, and explore the evolving behaviors of local celestial bodies by adopting higher precision observations, particularly for the once purely theoretical concept of wormholes or even white holes physically.

\section*{Acknowledgments}
We thank Professors Sergei. D. Odintsov and Bharat Ratra for beneficial feedback on astrophysics and cosmology. Deng Wang would like to thank Professor Jing-Ling Chen for helpful discussions about quantum information and quantum computation. This work is supported in part by the National Science Foundation of China.


\begin{thebibliography}{99}
\bibitem{1}
A. G. Riess et al., Observational evidence from supernovae for an accelerating universe and a cosmological constant, Astron. J. {\bf 116}, 1009 (1998).

\bibitem{2}
S. Perlmutter, M. S. Turner and M. White, Constraining dark energy with SNe Ia and large scale structure, Phys. Rev. Lett. {\bf 83}, 670 (1999).

\bibitem{3}
P. Ade et al., Planck 2015 results. XIII. Cosmological parameters, Astron. Astrophys. {\bf 594}, A13 (2016).

\bibitem{4}
S. Weinberg, The Cosmological Constant Problem, Rev. Mod. Phy. {\bf 61}, 1 (1989).

\bibitem{5}
E. Witten, Quantum gravity in de Sitter space, [arXiv: hep-th/0106109].

\bibitem{6}
R. R. Caldwell, A Phantom menace? Phys. Lett. B {\bf 545} 23--29 (2002).

\bibitem{7}
Y. Fujii, Origin of the gravitational constant and particle masses in a scale-invariant scalar-tensor theory, Phys. Rev. D {\bf26,} 2580 (1982).

\bibitem{8}
L. H. Ford, Cosmological-constant damping by unstable scalar fields, Phys. Rev. D {\bf35,} 2339 (1987).

\bibitem{9}
C. Wetterich, Cosmology and the Fate of Dilatation Symmetry, Nucl. Phys. B {\bf302,} 668 (1988).

\bibitem{10}
B. Ratra and P. J. E. Peebles, Cosmological consequences of a rolling homogeneous scalar field, Phys. Rev. D {\bf37,} 3406 (1988).

\bibitem{11}
S. M. Carroll, Quintessence and the Rest of the World: Suppressing Long-Range Interactions, Phys. Rev. Lett. {\bf81,} 3067 (1998).

\bibitem{12}
A. Hebecker, C. Witterich, Quintessential Adjustment of the Cosmological Constant, Phys. Rev. Lett. {\bf86,} 3339 (2000).

\bibitem{13}
A. Hebecker, C. Witterich, Natural quintessence? Phys. Lett. B {\bf497,} 281 (2001).

\bibitem{14}
R. R. Caldwell, M. Kamionkovski and N. N. Weinberg, Phantom energy and cosmic doomsday, Phys. Rev. Lett. {\bf 91}, 071301 (2003).

\bibitem{15}
P. Wang and X. Meng, Can vacuum decay in our universe? Class. Quant. Grav. {\bf 22,}  283-294 (2005).

\bibitem{16}
X. Meng, J. Ren and M. Hu, Friedmann cosmology with a generalized equation of state and bulk viscosity, Commun. Theor. Phys. {\bf47,} 379 (2007).

\bibitem{17}
J. Ren and X. Meng, Modified equation of state, scalar field and bulk viscosity in Friedmann universe, Phys. Lett. B {\bf636,} 5 (2006).


\bibitem{18}
J. Ren and X. Meng, Cosmological model with viscosity media (dark fluid) described by an effective equation of state, Phys. Lett. B {\bf633,} 1 (2006).

\bibitem{19}
M. Hu and X. Meng, Bulk viscous cosmology: statefinder and entropy, Phys. Lett. B {\bf635,} 186 (2006).

\bibitem{20}
X. Meng and X. Dou, Friedmann cosmology with bulk viscosity: a concrete model for dark energy, Commun. Theor. Phys. {\bf52,} 377 (2009).

\bibitem{21}
X. Dou and X. Meng, Bulk viscous cosmology: unified dark matter, Adv. Astron. {\bf1155,} 829340 (2011).

\bibitem{22}
A. Kamenshchik, U. Moschella and V. Pasquier, An Alternative to quintessence, Phys. Lett. B {\bf 511}, 265 (2001).

\bibitem{23}
S. Capozziello, Curvature quintessence, Int. J. Mod. Phys. D {\bf11,} 483 (2002).

\bibitem{24}
S. Capozziello et al., Curvature quintessence matched with observational data, Int. J. Mod. Phys. D {\bf12,} 1969 (2003).

\bibitem{25}
S. M. Carroll et al., Is cosmic speed - up due to new gravitational physics? Phys. Rev. D {\bf70,} 043528 (2004).

\bibitem{26}
S. Nojiri and Sergei D. Odintsov, Modified gravity with negative and positive powers of the curvature: Unification of the inflation and of the cosmic acceleration, Phys. Rev. D {\bf68,} 123512 (2003).

\bibitem{27}
S. Nojiri and Sergei D. Odintsov, Unified cosmic history in modified gravity: from F(R) theory to Lorentz non-invariant models, Phys. Rept. {\bf 505}, 59-144 (2011).

\bibitem{28}
S. Nojiri and Sergei D. Odintsov, Cosmological reconstruction of realistic modified F(R) gravities, Phys. Lett. B {\bf 681}, 74-80 (2009).

\bibitem{29}
J. P. Uzan, Cosmological scaling solutions of nonminimally coupled scalar fields, Phys. Rev. D {\bf59,} 123510 (1999).

\bibitem{30}
T. Chiba, Quintessence, The gravitational constant, and gravity, Phys. Rev. D {\bf60,} 083508 (1999).

\bibitem{31}
V. Sahni and A. A. Starobinsky, Reconstructing Dark Energy, Int. J. Mod. Phys. D {\bf 15,} 2105 (2006).

\bibitem{32}
P. Ruiz-Lapuente, Dark energy, gravitation and supernovae, Class. Quant. Grav. {\bf 24,} R91 (2007).

\bibitem{33}
L. Randall and R. Sundrum, A Large mass hierarchy from a small extra dimension, Phys. Rev. Lett. {\bf 83,} 3370 (1999).

\bibitem{34}
L. Randall and R. Sundrum, An Alternative to compactification, Phys. Rev. Lett. {\bf 83,} 4690 (1999).

\bibitem{35}
G. R. Davli, G. Gabadadze and M. Porrati, 4-D gravity on a brane in 5-D Minkowski space, Phys. Lett. B {\bf 485,} 208 (2000).

\bibitem{36}
T. Jacobson, Einstein-aether gravity: a status report, PoS QG-PH: 020 (2007).

\bibitem{37}
T. Jacobson, Extended Horava gravity and Einstein-aether theory, Phys. Rev. D {\bf 81}, 101502 (2010).

\bibitem{38}
Fernando Izaurieta et al., Standard General Relativity from Chern-Simons Gravity, Phys. Lett. B {\bf 678}, 213-217 (2009).


\bibitem{a2}
G. Dvali and G. Gabadadze, Gravity on a brane in infinite volume extra space, Phys. Rev. D {\bf 63}, 065007 (2001).

\bibitem{a3}
C. Deffayet, Cosmology on a brane in Minkowski bulk, Phys. Lett. B {\bf 502}, 199 (2001).

\bibitem{39}
D. Wang and X. Meng, Wormholes supported by phantom energy from Shan--Chen cosmological fluids, Eur. Phys. J. C {\bf 76}, 171 (2016).

\bibitem{40}
D. Wang and X. Meng, Modeling phantom energy wormholes from Shan--Chen fluids, [arXiv: 1512.03097].

\bibitem{41}
D. Wang and X. Meng, Traversable geometric dark energy wormholes constrained by astrophysical observations, Eur. Phys. J. C {\bf 76}, 484 (2016).

\bibitem{42}
D. Wang and X. Meng, Traversable holographic dark energy wormholes constrained by astronomical observations, [arXiv: 1602.04699].

\bibitem{43}
L. Flamm, Beitr¨¢ge zur Einsteinschen Gravitationstheorie, Phys. Z. {\bf 17}, 448 (1916).

\bibitem{44}
A. Einstein and N. Rosen, The Particle Problem in the General Theory of Relativity, Phys. Rev. {\bf 48}, 73-77 (1935).

\bibitem{45}
J. A. Wheeler, Geons, Phys. Rev. {\bf 97}, 511-536 (1955).

\bibitem{46}
C. W. Misner and J. A. Wheeler, Classical physics as geometry: Gravitation, electromagnetism, unquantized charge, and mass as properties of curved empty space, Annals Phys. {\bf 2}, 525 (1957).

\bibitem{47}
K. A. Bronnikov, Scalar-tensor theory and scalar charge, Acta Phys. Pol. B {\bf 4}, 251 (1973).

\bibitem{48}
H. G. Ellis, Ether flow through a drainhole - a particle model in general relativity, J. Math. Phys. {\bf 14}, 104 (1973).

\bibitem{49}
H. G. Ellis, The Evolving, Flowless Drain Hole: A Nongravitating Particle Model In General Relativity Theory, Gen. Rel. Grav. {\bf 10}, 105--123 (1979).

\bibitem{50}
M. S. Moris and K. S. Thorne, Wormholes in space-time and their use for interstellar travel: A tool for teaching general relativity, Am. J. Phys. {\bf 56,} 395 (1988).

\bibitem{51}
M. Visser, Traversable wormholes: Some simple examples, Phys. Rev. D {\bf 39}, 3182 (1989).

\bibitem{52}
M. Visser, Traversable wormholes from surgically modified Schwarzschild space-times, Nucl. Phys. B {\bf 328}, 203 (1989).

\bibitem{53}
M. Visser, Quantum Mechanical Stabilization of Minkowski Signature Wormholes, Phys. Lett. B {\bf 242}, 24 (1990).

\bibitem{54}
E. Poisson and M. Visser, Thin shell wormholes: Linearization stability, Phys. Rev. D {\bf 52}, 7318 (1995).

\bibitem{55}
S. V. Sushkov, Wormholes supported by a phantom energy, Phys. Rev. D {\bf 71}, 043520 (2005).

\bibitem{56}
F. S. N. Lobo, Phantom energy traversable wormholes, Phys. Rev. D {\bf 71}, 084011 (2005).

\bibitem{57}
D. Wang and X. Meng, Observational constraints and diagnostics for time-dependent dark energy models, [arXiv: 1603.00699].


\bibitem{58}
D. Wang and X. Meng, Observational constraints and differential diagnosis for cosmic evolutionary models, [arXiv: 1603.08112].

\bibitem{59}
D. Wang and X. Meng, Reconstructing f(R) gravity from viscous cosmology constrained by observations, [arXiv: 1604.02951].

\bibitem{60}
C. Deffayet et al., Accelerated universe from gravity leaking to extra dimensions, Phys. Rev. D {\bf 65}, 044023 (2002).

\bibitem{61}
B. P. Abbott et al, Observation of Gravitational Waves from a Binary Black Hole Merger, Phys. Rev. Lett. {\bf 116}, 061102 (2016).





































\end{thebibliography}
\end{document}